\documentclass[iop,apj,numberedappendix]{emulateapj}
\usepackage{graphicx}
\usepackage{apjfonts}
\usepackage{array,color}
\usepackage{amsmath}
\usepackage{amssymb}
\usepackage{amsfonts}
\usepackage{amsbsy}
\usepackage{latexsym}
\usepackage{dcolumn}% Align table columns on decimal point
\usepackage{bm}% bold math
\usepackage{epsf}
\newcommand{\be}{\begin{equation}}
\newcommand{\ee}{\end{equation}}
\newcommand{\ba}{\begin{eqnarray}}
\newcommand{\ea}{\end{eqnarray}}
\newcommand{\Mpc}{$\;h^{-1}\;\textrm{Mpc}\;$}
\newcommand{\Msolar}{$\;h^{-1}\;\rm{M_\odot}\;$}
\newcommand{\hMpc}{$\;h\;\textrm{Mpc}^{-1}\;$}
\newcommand{\bi}{\begin{itemize}}
\newcommand{\ei}{\end{itemize}}
\newcommand{\bfi}{\begin{figure}
\epsfxsize=9cm 
\epsffile}
\newcommand{\efi}{\end{figure}}

\begin{document}
\title{The thermal SZ tomography}
\author{Jiawei Shao$^{1,2}$,Pengjie Zhang$^1$, Weipeng Lin$^1$, Yipeng Jing$^1$}
\affil{$^1$Key Laboratory for Research in Galaxies and Cosmology, Shanghai
  Astronomical Observatory, Nandan Road 80, Shanghai, 200030, China
\\$^2$Graduate School of the Chinese Academy of Sciences, 19A, Yuquan Road,
  Beijing, China}
\email{jwshao@shao.ac.cn}
\begin{abstract}
	The thermal Sunyaev-Zel'dovich (tSZ) effect directly measures the
thermal pressure of free  electrons integrated along the line of sight and thus
contains valuable information on the thermal history of the universe. However,
the redshift information is entangled in the projection along  the line of
sight. This projection effect severely degrades the power of the tSZ effect to
reconstruct the thermal history. We investigate the tSZ tomography technique to
recover  this otherwise lost redshift information by cross correlating the tSZ
effect with galaxies of known redshifts, or alternatively with matter
distribution reconstructed from weak lensing tomography. We investigate in
detail the 3D distribution of the gas thermal pressure and its relation with the
matter distribution, through our adiabatic hydrodynamic simulation and the one
with additional gastrophysics including radiative cooling, star formation and
supernova feedback. (1) We find a strong correlation between the gas pressure
and matter distribution, with a typical cross correlation coefficient $r\ga 0.7$
at $k\la 3h/$Mpc and $z<2$. This tight correlation will enable robust cross
correlation measurement between SZ surveys such as Planck, ACT and SPT and
lensing surveys such as DES and LSST, at $\ga 20$-$100 \sigma$ level. (2) We
propose a tomography technique to convert the measured cross correlation into the
contribution from gas in each redshift bin to the tSZ power spectrum.
Uncertainties in  gastrophysics may affect the reconstruction at $\sim 2\%$
level, due to the $\sim1\%$ impact of  gastrophysics on  $r$, found in our
simulations. However, we find that the same gastrophysics affects the tSZ power
spectrum at $\sim 40\%$ level, so it is robust to infer the gastrophysics from
the reconstructed redshift resolved contribution.  
\end{abstract}


\section{Introduction}
Free electrons in the universe reveal their existence in the CMB sky through 
their inverse Compton scattering  of CMB photons. The induced secondary CMB 
temperature anisotropies, proportional to the electron thermal energy
integrated  along the line of sight, are the well known thermal Sunyaev-
Zel'dovich (tSZ) effect \citep{SZ1972,SZ1980}. Since massive galaxy clusters contain
large reservoir of hot electrons, the generated tSZ effect can thus overwhelm
the primary CMB around  cluster scales. For this reason, the tSZ effect
of dozens of galaxy clusters has been measured by various  experiments (refer
to \citealt{Carlstrom02,Reese02,Jones05,LaRoque06,Bonamente06} for
reviews). On the other hand, blindly detecting the tSZ effect in random
directions of sky is much more difficult, since the expected signal is
overwhelmed by the primary CMB fluctuations at large angular scales and heavily
polluted by astrophysical foregrounds at smaller angular scales. This causes
ambiguity in interpreting the observed small scale CMB power excess
\citep{Dawson02, Dawson06, Mason03, Goldstein03, Runyan03, Kuo04,
Readhead04, Bond05, Reichardt09, Sharp10}. However, the situation is improving
significantly by 
ongoing and proposed ground surveys such as  
SZA\footnote{http://astro.uchicago.edu/sza/},
ACT\footnote{http://www.physics.princeton.edu/act/},
APEX\footnote{http://bolo.berkeley.edu/apexsz/},
SPT\footnote{http://pole.uchicago.edu/} and the Planck satellite. Recently the
SPT collaboration reported the first SZ power spectrum measurement at $\ell
\sim 3000$ \citep{Lueker2010}, and the ACT collaboration reported their SZ power
spectrum analysis at similar scales \citep{Fowler2010, Dunkley2010} and found
consistent results. These tSZ power spectrum measurements will be further
improved by orders of magnitude improvement in sky coverage and better analysis
enabled by more frequency bands. 

Precision mapping of the SZ sky is of great importance to both cosmology and
astrophysics. The SZ effect is a powerful finder of galaxy clusters at high
redshifts. The efficiency of free electrons to generate the thermal SZ effect
is redshift independent. Photons originated from redshift $z$ suffer a factor
of $1+z$ energy loss caused by the cosmic expansion. On the other hand,   CMB
photons that electrons scatter at redshift $z$ are a factor of $1+z$ more
energetic than CMB photons today. These two effects cancel out exactly and enables
galaxy clusters to be detected at high redshift without extra effort, opposite
to the X-ray cluster finding. An exciting advance in this area is the first
discovery of 3 new galaxy clusters through the tSZ effect by the SPT group
\citep{Staniszewski2009}.  By far, a few dozen galaxy clusters have been
detected through blind SZ surveys done by SPT \citep{Vanderlinde2010}
and ACT \citep{Menanteau2010}.

The tSZ effect is also a powerful probe to the thermal history of the universe,
since it directly probes the thermal energy of intergalactic medium (IGM) and
intracluster medium (ICM). Numerical simulations show that, the amplitude of
the SZ signal is sensitive to the amount of radiative cooling and energy
feedback \citep{SWH01,daSilva01,White02,Lin04}. However, it is not
straightforward to extract information on these astrophysical processes from
the tSZ measurements alone. First of all, research shows that  there exist
great degeneracies between different competing processes. Even worse, the tSZ
effect only measures the electron thermal energy projected along the line of
sight.  The redshift  information of these astrophysical processes is thus
entangled in  the projection. 

\citet{Zhang01} proposed to recover the redshift information by cross
correlating the tSZ effect with galaxies with at least photo-z information. The
idea is that galaxies in a given redshift bin should strongly correlate
with the  tSZ signal from the same redshift bin. A key link between the
measured cross correlation and the gas pressure auto-correlation that we
want to extract is the cross correlation coefficient $r$ between the thermal
energy and the galaxy number density.  Assuming a constant $r$, the time
resolved thermal energy distribution can be reconstructed self consistently.
This SZ tomography technique would be applicable in reality, since SZ surveys
often have follow-ups of galaxy  surveys. For example, the dark energy
survey\footnote{http://www.darkenergysurvey.org/}  will cover the SPT sky and
measure photometric redshifts of $\sim 10^8$ galaxies up to $z=1.3$.   The
lensing tomography also helps to reconstruct the 3D matter distribution, which
can also be  correlated with the SZ map to make the SZ tomography.

In the current paper, we reformulate this SZ tomography technique and explore
the possibility to improve its robustness.  In the main text, we will choose
the matter distribution reconstructed from weak lensing tomography as the
tracer of the large scale structure to investigate the tSZ tomography.  We no
longer approximate $r$ as a constant. Rather, we rely on numerical simulations
to quantify its scale and redshift dependence. We are able to show that, $r$ is
insensitive to gastrophysics such as radiative cooling and supernova (SN)
feedback. Namely, $r$ calculated from adiabatic hydrodynamic simulations should
be sufficiently accurate for the purpose of the tSZ tomography, even with the
presence of radiative cooling and SN 
feedback. We are then able to take this $r$ as input to perform the SZ
tomography. In the appendix \ref{app:galaxy}, we will investigate the
feasibility to perform the tSZ tomography by cross correlating with galaxy distribution. 

The paper is organized as follows. In \S \ref{sec:tomography}, we introduce the
tSZ effect and explain the SZ tomography technique. We then test its feasibility
against our high precision hydrodynamic simulations (\S \ref{sec:simulation}).
We also quantify the dependence of various tSZ statistics on the extra
gastrophysics. Although most
quantities are sensitive to these gastrophysical processes, we find that the
cross correlation coefficient $r$ only weakly depends on them (\S
\ref{subsec:r}). This feature allows us to use $r$ calculated from the
adiabatic simulations as the input of the tSZ tomography, despite the
existence of complicated gastrophysics in the real universe. We discuss
and make conclusion in \S \ref{sec:discussion}. In the appendix, we forecast
the accuracy of performing the  tSZ tomography with lensing surveys
(\S \ref{app:lensing}) and  discuss complexities with galaxy redshift surveys
(\S \ref{app:galaxy}). 

\section{The thermal SZ effect and the SZ tomography}
\label{sec:tomography}
The tSZ effect induces a new source of CMB temperature fluctuations with the
amplitude
\begin{equation}
	\frac{\Delta T(\boldsymbol{\theta})} {T_{CMB}}
		=g(x) y(\boldsymbol{\theta})\ .
\end{equation}
Here, $\boldsymbol{\theta}$ is the direction on the sky. $g(x)$ describes the
spectral dependence. In the non-relativistic limit, 
\be
	g(x)=\left( x \frac{e^x+1}{e^x-1} - 4 \right)\ ,
\ee
where $x\equiv h \nu /k_B T_{CMB} = \nu/ 56.84 $ GHz and $\nu$ is the observed
frequency of CMB photons. The  Comptonization parameter $y(\theta)$ is  
\be
y = \frac{ \sigma_T}{m_e c^2} \int {a d\chi n_e k_B T_e},
\label{y}
\ee
where $n_e k_B T_e$ is the hot electron pressure. $\chi$, $n_e$, $k_B$, $T_e$
and $\sigma_T$ are  the comoving diameter distance, number density of free
electrons, the Boltzmann constant, electron   temperature and the Thompson
scattering cross section respectively.  

 The spectral dependence of the tSZ effect is unique. The tSZ effect shows as
 CMB temperature decrements at $\nu<218$ GHz and as increments at $\nu>218$
 GHz. The spectral function $g(x)\rightarrow -2$ at the Rayleigh-Jeans band
 $x\ll 1$ and $g(x)\rightarrow x-4$ at $x\gg 1$.  This unique spectral
 dependence allows a clean separation of the tSZ effect from other CMB
 components in multi-band CMB surveys. 

The $y$ parameter contains key information on the thermal history of the
universe. However, since it only  measures the projected electron thermal
energy along the line of sight, the redshift information is smeared 
by this projection effect. Our SZ tomography technique aims to recover the
otherwise lost redshift information in the tSZ effect. 

\subsection{The tSZ tomography}
One of the most widely used statistical quantities of the tSZ effect is the
angular power spectrum $C^{\rm tSZ}_l$. Throughout this paper, unless otherwise
specified, we will focus on the Rayleigh-Jeans limit $\Delta
T/T=-2y$ (however in Appendix \ref{app:lensing}, we set $g= -0.95$
in order to forecast the measurement error for ACT and SPT at $\nu=150$GHz).
Under the Limber approximation \citep{Limber1954}, $C^{\rm tSZ}_l$ is related
to the 3D thermal pressure power spectrum $\Delta^2_P(k,z)$ by the following
relation,
\begin{equation}
\label{eqn:SZcl}
\frac{l^2}{2\pi}C_l^{\rm tSZ}=\int_0^{\chi_{\rm CMB}}
\Delta^2_P(k=\frac{l}{\chi},z)  W_{\rm tSZ}^2(z) \chi d\chi\ . 
\end{equation}
We have adopted the flat cosmology in the above expression. The weighting
function is
\be
W_{\rm tSZ}(z)=-2 \sigma_T a\frac{\langle n_ek_BT_e\rangle}{m_ec^2}\ .
\ee
We here express the power spectrum in dimensionless form as the variance
per $\ln k$, $\Delta_P^2(k)=k^3 P_P(k)/2\pi^2$, where the dimensional power
spectrum $P_P(k)= \left< \delta_P({\bf k}) \delta^*_P({\bf
k})\right>$, and $\delta_P(k)$ is the Fourier transform of the fractional
thermal pressure fluctuations $\delta_P\equiv n_ek_BT_e/\langle
n_ek_BT_e\rangle-1$. Following similar notation, we can define the dimensional matter power
spectrum $P_m=\left< \delta_m({\bf k})\delta_m^*({\bf k})\right>$ and the
dimensional cross power spectrum $P_{\rm Pm}=\left<\delta_m({\bf
k})\delta_P^*({\bf
k})\right>$, thus $\Delta^2_m$ and $\Delta^2_{\rm Pm}$ accordingly.

Our SZ tomography technique aims to reconstruct the time resolved
$\Delta^2_P(k,z)  W_{\rm tSZ}^2(z)$. This quantity tells us the overall
amplitude of the thermal energy and the clustering strength at redshift $z$.
The starting point of this technique is that, tSZ effect is correlated with the
underlying matter distribution. Theoretically, given the redshift information
of the underlying matter distribution, we can manage to recover the redshift
information of the tSZ effect. The original tomography is presented in the
variation formalism \citep{Zhang01}, while in the current paper, we reformulate
it in a more straightforward manner. The key idea of the SZ tomography is that,
dark matter distributed in a certain redshift range correlates with the
tSZ signals contributed by the IGM in the {\it same} redshift range. In
practice, \citet{Massey07} proposed a mass reconstruction method using lensing
tomography method. Therefore, if we have the dark matter distribution with
redshift information, we are able to carry out SZ tomography in the
corresponding redshift bins. 

Let's consider matter distribution in redshift range from $z=0$ to $z=z_s$. We
can divide this redshift range into redshift bins, in which a typical redshift
bin is $z_i-\Delta z/2\le z \le z_i+\Delta z/2$.  In this narrow redshift bin,
we can then define a weighted surface matter density
\begin{equation}
 \Sigma_m=\int_{\chi_i+\frac{\Delta \chi_i}{2}}^{\chi_i+\frac{\Delta \chi_i}{2}} \delta_m
W_m(\chi) d\chi \ , 
\end{equation}
with a proper weighting function $W_m$. Here, $\chi_i\equiv \chi(z_i)$ and
$\Delta \chi_i=\chi(z_i+\Delta z_i/2)-\chi(z_i-\Delta z_i/2)$. The weak lensing tomography technique
enables us to reconstruct a map of $\Sigma_m$ in a lensing survey with source
redshift information and $W_m$ is
defined accordingly. Lensing tomography has the potential to reconstruct maps
of width of $\Delta z\sim 0.1$, in which $W_m$ can be treated as a
constant. In case of tSZ-galaxy cross correlation, $\Sigma_m$ is replaced by
$\Sigma_G$ and $W_m$ is replaced by $W_G$. Without loss of generality,
throughout this paper,  we will
treat $\Sigma_m$ as observable. 

%{\bf In the situation of weak lensing, $W_m(\chi)=3\Omega_0
%H_0^2\chi(1-\chi/\chi_s)/2a$, where $\chi_s$ is the comoving distance of the
%source galaxy.}

To extract the tSZ redshift information, we cross correlate  tSZ maps with
$\Sigma_m$ of overlapping sky. At sub-degree scales of interest, the Limber
approximation holds. Under this limit, the cross correlation signal is and is
only  contributed by matter and gas pressure in this very redshift bin. We can
thus express the tSZ-matter cross power spectrum with the Limber integral, 
\begin{equation}
\label{eqn:cl_cross2}
\frac{l^2}{2\pi}C^{\rm tSZ-m}_l=\int_{\chi_i-\frac{\Delta
\chi_i}{2}}^{\chi_i+\frac{\Delta \chi_i}{2}}
\Delta^2_{Pm}(k=\frac{l}{\chi},\chi)  W_{\rm tSZ}(\chi) W_m(\chi) \chi d\chi \
.
\end{equation}

For a narrow redshift bin, all the functions in the integrand vary
slowly across this redshift bin. We thus 
have an approximation
\begin{eqnarray}
\label{eqn:cl_cross}
\frac{l^2}{2\pi}C^{\rm tSZ-m}_l&\simeq 
\Delta^2_{Pm}(k=\frac{l}{\chi_i},\chi_i) W_{\rm  tSZ}(\chi_i) W_m(\chi_i)
\chi_i \Delta \chi_i\ .
\end{eqnarray}
 It is clear from the above equation that the
cross correlation 
between the thermal SZ maps with matter distribution picks out and, to a good
approximation, only picks out relevant information within the given redshift
bin. This is a key step to recover the redshift information of the thermal SZ
effect. 

For similar argument, the auto angular power spectrum of matter in the same
redshift bin is
\begin{equation}\begin{split}
\label{eqn:cl_gal}
\frac{l^2}{2\pi}C^{m}_l&=\int_{\chi_i-\frac{\Delta
\chi}{2}}^{\chi_i+\frac{\Delta \chi}{2}}
\Delta^2_{m}(k=\frac{l}{\chi},\chi) W_{m}^2 \chi d\chi \\
&\simeq \Delta^2_m(k=\frac{l}{\chi_i},\chi_i) W_{m}^2(\chi_i) \chi_i
\Delta \chi_i
\end{split}\end{equation}

The next step is to recover the thermal SZ contribution in the given redshift
bin with the correct weighting, namely, $\Delta^2_P(k,z) W_{\rm tSZ}^2(z)$.
This step requires a key input, namely the cross correlation coefficient $r$
between fluctuations in the gas pressure and the dark matter distribution,
defined by 
\ba
\label{eqn:corr_definition}
	r(k,z)=\frac{\Delta^2_{\rm Pm}(k,z)}{\Delta_P(k,z) \Delta_m(k,z)}\ .
\ea
Combining Eq. \ref{eqn:cl_cross} and Eq. \ref{eqn:cl_gal}, we obtain {\it the
redshift resolved gas pressure auto-correlation} (with correct weighting)
 \begin{equation}
\begin{split}
\label{eqn:tomography}
\Delta^2_P(k,z_i)& W_{\rm tSZ}^2(z_i)=r^{-2}(k,z_i)
\frac{[\Delta^2_{\rm Pm}(k,z_i) W_{\rm tSZ}]^2}{\Delta^2_m(k,z_i)}\\
	&\simeq r^{-2}(k,z_i)\frac{l^2\left[C_l^{\rm
      tSZ-m}\right]^2}{ 2\pi C_l^{m}} 
 \frac{1}{\chi_i \Delta \chi_i}\ .
\end{split}
\end{equation}
In this relation, the angular power spectra $C_l^{\rm
tSZ-m}$ and $C_l^m$ are evaluated at $l=k\chi_i$.

To better understand the meaning of the above equation, we remind that
Eq. \ref{eqn:SZcl} can be expressed as 
\be
\label{eqn:tSZsum}
\frac{l^2C_l^{\rm tSZ}}{2\pi}\simeq \sum_i \left[\Delta^2_P(k=\frac{l}{\chi_i},z_i)W_{\rm
    tSZ}^2(z_i)\right]\chi_i\Delta \chi_i \ . 
\ee
Hence Eq. \ref{eqn:tomography} indeed tells us the contribution to the tSZ
power spectrum from the $i$-th redshift bin.

Given dark matter distribution with redshift information, the tSZ angular power
spectra can be measured directly by combining the SZ surveys and large scale
structure surveys. The proper weighting function $W_m$ is given by large scale
structure survey, and $\chi$ can be calculated given the cosmology model. As
long as $r(k,z)$ can be figured out, one can apply this equation to obtain the
redshift resolved gas pressure auto correlation. Large scale structure surveys
are usually limited to $z\la 2$, so the tSZ tomography and the redshift resolved
auto correlation reconstruction are limited to $z\la 2$ too. However, $C^{\rm
  tSZ}_l$ is also an observable, which contains contribution from all redshifts.
Subtracting the $z\la 2$ contribution inferred from our tSZ tomography, we can
infer the tSZ contribution from higher redshifts, such as the one from supernova
explosion of first stars \citep{Oh03}. 

We summarize the thermal SZ tomography using lensing tomography method as
follows.
\bi
\item Reconstruct the 3D matter distribution from large scale structure surveys.
\item Measure the cross power spectrum $C_l^{\rm tSZ-m}$ between the thermal SZ
effect and the reconstructed matter in a given redshift bin. 
\item Measure the matter angular power spectrum $C_l^{m}$ in this redshift bin. 
\item Reconstruct the tSZ contribution from this redshift bin to the
  tSZ power spectrum through
  Eq. \ref{eqn:tomography}.
\ei

A key input required in this SZ tomography is the quantity $r$. 
\citet{Zhang01} adopted a simplification that $r$ is a constant. They further
showed that, the value of $r$ can be measured combining
Eq.\ref{eqn:tomography} and the measured SZ power spectrum. This approach is
self consistent and able to provide a quick realization of the SZ
tomography. However, improvements must be made to do precision SZ
tomography. As shown in \S \ref{subsec:r}, the assumption that $r$ is a
constant is only accurate at the level of $\sim 20\%$. Furthermore, to obtain
the value of $r$ unbiasedly from the SZ power spectrum, the galaxy surveys must
well cover the SZ redshift range ($z\la 2$). This is challenging for galaxy
surveys.  

Since $r(k,z)$ is such a key quantity in the SZ tomography, a natural question
arises: {\it Can one robustly predict $r$?} This seems challenging, given
the fact that various complicated gastrophysical processes affect the tSZ
effect. Surprisingly, robust prediction on $r$ is likely feasible. The key
point is that, these gastrophysics affects $\Delta^2_P$ and $\Delta^2_{Pm}$ in
basically the same way, so that these effects roughly cancel out in $r$ by
definition. To quantify the dependence of $r$ on these gastrophysics, we compare
two sets of hydrodynamic simulations with and without radiative cooling, star
formation and SN feedback. These simulations confirm the above naive speculation
and find that, these additional gastrophysics can only affect $r$ at $1\%$
level, despite the fact that they alter the SZ power spectrum by $\sim 40\%$. 

\section{Simulations}
\label{sec:simulation}
The tSZ statistics has been studied by both semi-analytical models
\citep{Cooray00,Zhang01,KS02,Zhang03,Zhang04,Zhang07} and numerical simulations
\citep{Persi95,Refregier00,Refregier02,daSilva00,daSilva01,SBP01,SWH01,White02,Lin04,Zhang04,
Cao2007,2007MNRAS.378.1259R,Hallman2007,Hallman2009}.  Beyond the gravitational
heating mechanism, some works
\citep{daSilva01,SWH01,White02,Lin04,2007MNRAS.378.1259R,Scannapieco08} have
incorporated additional gastrophysics such as radiative cooling, preheating,
SN/AGN feedback. These studies found that these processes suppress the small
scale SZ power spectrum. But none of them addressed the dependence of the cross
correlation coefficient $r$ on these processes, whose investigation is the key
goal of this paper. 

In this paper, we analyze a controlled set of hydrodynamic simulations for the
relevant SZ statistics.  The cosmology adopted is a $\Lambda{\rm CDM}$
cosmology:   
$\Omega_\Lambda = 0.732, \Omega_0=0.268, \Omega_b=0.04448, h=0.71,
\sigma_8=0.85$.  The simulations are run by  GADGET2 code
\citep{Springel05}, with $100$ \Mpc box size, and $512^3$ particles for both
dark matter and SPH particles. The masses of the dark
matter and SPH particles are $4.62 \times  10^8 $ \Msolar and $9.20 \times
10^7$ \Msolar respectively. One simulation only includes gravitational heating
and the other includes radiative cooling and star formation with SN feedback.
We  refer these two as adiabatic and non-adiabatic run, respectively. For the
non-adiabatic run, $7.63\%$ of  gas particles turn into stars by $z=0$,
resulting in stellar mass density $\Omega_*=0.0034$. Both simulations start
from redshift $z=120$.  We output 60 snapshots, which are equally spaced in
$\ln{a}$. Refer to \citet{Lin2006,Jing06} for more details of the simulations.

We derive the temperature $T_i$ of the $i$-th gas particle as follows
\begin{equation}
	k_B T_i=(\gamma-1) \mu m_p u_i,
\end{equation}
with $\mu=0.588$ the mean molecule weight, $\gamma=5/3$ the specific heat
ratio for monatomic gas, and $u_i$ the internal energy per unit mass of the
$i$-th particle. When the temperature of a SPH particle is higher than $10000$
K, we consider it as ionized, otherwise as neutral. We sum up all the 
SPH particles to calculate the density-weighted temperature $T_\rho$
at all the available snapshots. From these measurements, we are then able to
calculate the $y$ parameter \S \ref{subsec:y}.

We use the fast Fourier transform (FFT) to calculate the power spectra of
dark matter, gas pressure and the cross power spectrum between the two.
For dark matter particles, we assign the masses to $512^3$ grids using the
Clouds-In-Cells (CIC) algorithm\citep{Hockney1981, Jing05}. On the other hand,
we assign the thermal energy of SPH particles to the same grids using the SPH
kernel while assuring the energy conservation. The gas pressure on the grid is
converted from the gas thermal energy assigned. We remove the shot noises from
the corresponding raw power spectra. However, since we are mainly interested
in scales well below the Nyquist frequency (e.g. $k\la 16 $\hMpc), we don't
correct for the alias effect.\footnote{The alias effect would affect the
power spectra to scales as large as a quarter of the Nyquist frequency, namely
$k\sim 4$\hMpc in our simulations \citep{Jing05}. The methods proposed by
\citet{Jing05} and \citet{Cui2008} can effectively correct for this effect.
However, to the first order approximation, it affects the corresponding power
spectra in the two simulations in the same manner. Since we're mainly concerned
about the differences between the two power spectra, we neglect this effect in
this paper.} Results are shown in \S\ref{subsec:p} \& \ref{subsec:r}. 

\bfi{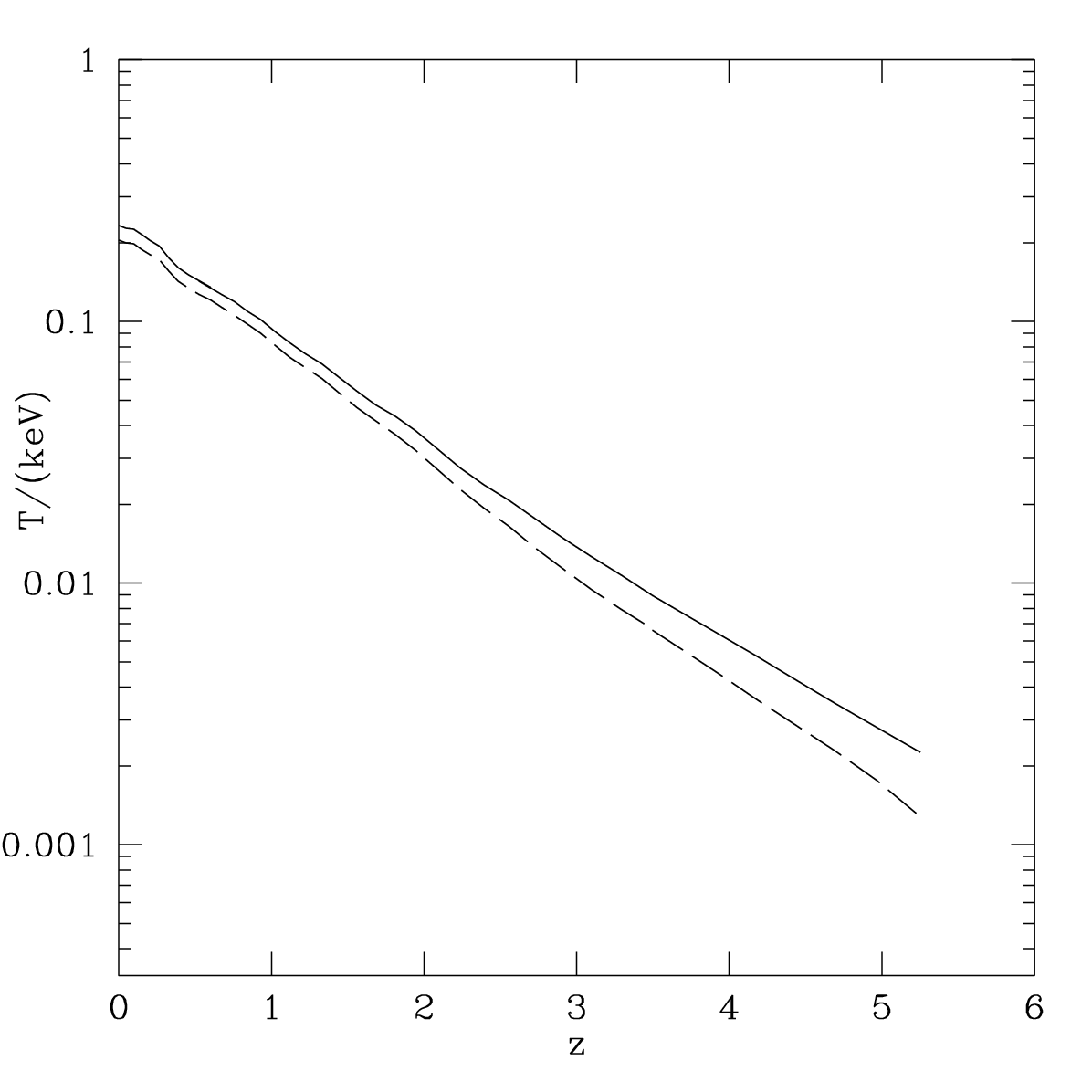}
\caption{The redshift evolution of the density-weighted temperature $T_\rho$ in
the adiabatic (solid line) and the non-adiabatic simulation (dashed line)
with cooling, star formation and SN feedback.}
\label{tav}
\efi

\subsection{The tSZ mean $y$ parameter and the gas density weighted
temperature} 
\label{subsec:y}
	The mean Comptonization $y$ parameter is given by the following equation
\be
\label{eqn:meany}
\bar{y}=\frac{\sigma_T}{m_ec^2}\int \bar{n}_ek_B T_{\rho} a d\chi .
\ee
Here $T_\rho\equiv \langle n_eT\rangle/\bar{n}_e$ is the density weighted
temperature, which is computed by the following equivalent expression
\begin{equation}
	T_\rho=\frac{\sum_i T_i m_i}{\sum_i m_i}\ \ .
\end{equation}
The sum is over all gas particles, assuming gas with $T_i \ge 10000$
K as ionized, otherwise neutral.

\bfi{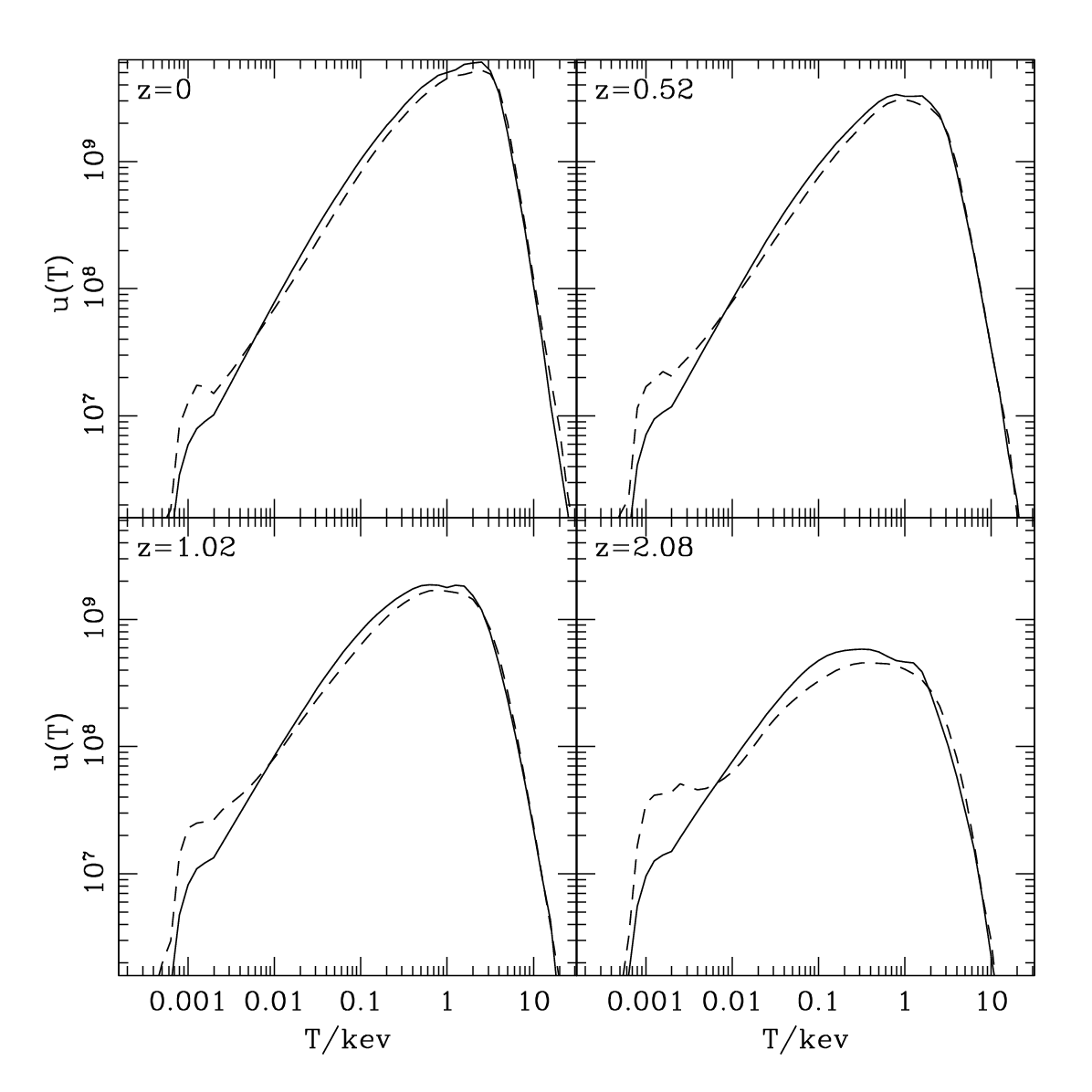}
\caption{The distribution of internal energy at different temperature $u(T)$
which is defined as $\int u(T) d\ln T = U$, where $U$ is the total internal
energy.  The additional processes brings the total internal energy down by
12.0\%, 10.9\%, 11.5\% and 16.6\% at z=0, 0,52, 1.02 and 2.08 respectively.
The same arbitrary unit is adopted for both the simulations. For both the
adiabatic (solid lines) and non-adiabatic (dashed lines) simulation, the most
contributions come from the gas at temperature around $1-2$ keV. And the lower
the redshift, the more contributions from higher temperature components. 
}
\label{fracu}
\efi

Figure \ref{tav} shows the redshift evolution of the density weighted temperature
for the adiabatic (solid lines) and the non-adiabatic (dashed lines)
simulation. As shown, the additional gastrophysical processes
suppress the mass weighted temperature $T_\rho$, which confirms the
earlier results by \citet{White02}, though more evident than theirs at z=0.

We further look into the effect of cooling and feedback in redistributing the
energy budget into regions with different temperatures.  The logarithmic
contribution to the total thermal energy from regions with different
temperatures is shown in Fig. \ref{fracu}.  We can see clearly that, for the 4
presented redshifts, the most contributions of the total energy come from the
regions with $T\sim 1-2$ keV at lower redshifts except $z=2.08$, for both
adiabatic and non-adiabatic simulations. The figure also shows that gas with
temperature higher than $\sim 2$ keV is slightly affected by these additional
gastrophysics. Most of these gas
lies in clusters, which have deep gravitational potential wells to offset the
feedback effect. Also, the star formation rate is relatively low in cluster
regions, resulting in less feedback intensity and also less depletion of hot
gas by cooling. On the other hand, gas cooler than $\sim 1$ keV has opposite
fate. Most of these gas lies in galaxy groups or in IGM , susceptible
to cooling and feedback. This suppresses their contribution to the total
energy. The combined effect is that, the total thermal
energy is suppressed by a factor of 12.0\%, 10.9\%, 11.5\% and 16.6\% at z=0,
0,52, 1.02 and 2.08 respectively. 

Given $T_\rho$ measured at the 60 simulation snapshots, we are able to
integrate it over the whole relevant redshift range to calculate $y$, according
to Eq. \ref{eqn:meany}. We adopt a simple linear, while stable, interpolation
to model $T_{\rho}$ at redshift falling  between two adjacent output redshifts.
For both cases, about a half of the $y$ signal comes from $z<1$. We obtain
$y=2.07\times 10^{-6}$ for the adiabatic run and $y=1.77\times 10^{-6}$ for the
non-adiabatic run. These values of $y$ are lower than the results in e.g.
\citep{daSilva01, White02,Zhang04}, so as the result of $T_{\rho}$. On the
other hand, it is higher than some recent simulation results, such as that
of \citet{2007MNRAS.378.1259R} using GADGET2. Differences in cosmological
parameters, such as difference in $\sigma_8$ and $\Omega_b$, can only account
for part of the discrepancy, while cosmic variance, differences in the
input gastrophysics and numerical codes may be responsible for the remains. We
postpone further investigation of this issue elsewhere.\footnote{We thank
Volker Springel for providing the result of an independent GADGET2 simulation,
which allows us to do a preliminary check.  The cosmological parameters of this
simulation are identical to ours, except the different initial conditions and
$\sigma_8=0.9$. This simulation gives $T_{\rho}=0.283$ keV. Taking the scaling
relation $T_{\rho}\propto \sigma_8^{3.05-0.15\Omega_0}$ \citep{Zhang04}, we
estimate that this simulation would result in $T_{\rho}=0.243$ keV for
$\sigma_8=0.85$, in good agreement with ours.}

\bfi{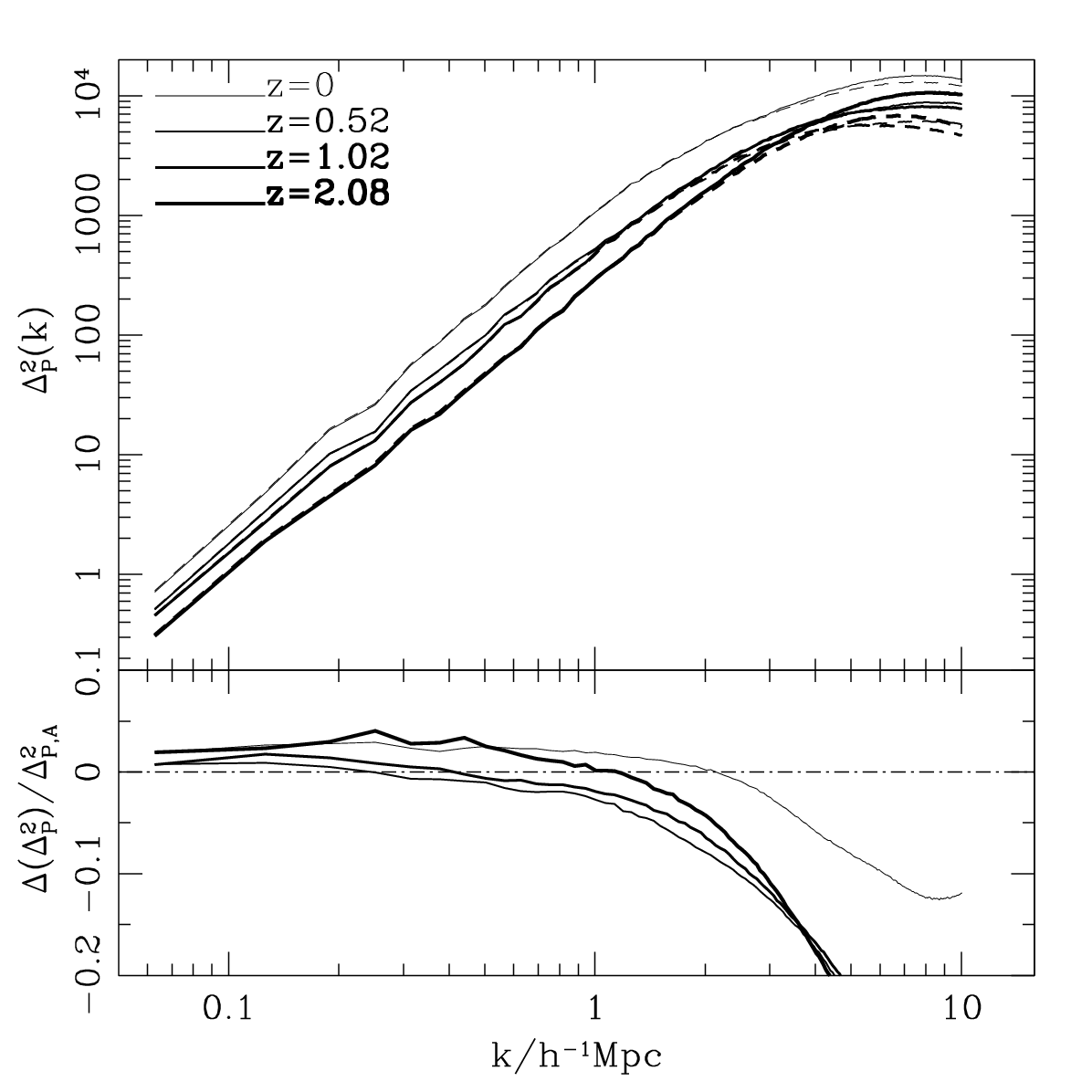}
\caption{\textbf{Top panel}: The power spectra of gas pressure $\Delta^2_P(k)$
at redshifts z=0,0.52,1.02 and 2.08. At small scales, the pressure power
spectra of non-adiabatic simulation (dashed lines) are greatly reduced by
cooling and SN feedback, even more significant to higher redshifts, compared
to the adiabatic one (solid lines). However, the powers at large scales are
slightly boosted.
\textbf{Bottom panel}: The differences between the power spectra defined by
$\Delta(\Delta^2_P)/\Delta^2_{P,A} = (\Delta^2_{P,NA}(k)-\Delta^2_{P,A}(k))/\Delta^2_{P,A}(k)$} 
\label{ps}
\efi

\subsection{The 3D gas pressure power spectrum and the tSZ power spectrum}
\label{subsec:p}
Most of the simulated SZ power spectra in the literature are directly calculated
from the simulated 2D SZ maps. In this paper we take an alternative approach, as
adopted by \citet{Refregier00,Refregier02}. For each simulation output, we can
calculate the 3D power spectrum of the gas thermal pressure $\Delta^2_P$. The
2D SZ power spectrum is obtained through Eq. (\ref{eqn:SZcl}) and we
interpolate linearly between adjacent snapshots to model $\Delta^2_p(k)$ at any
other redshifts.

$\Delta^2_P$ at $z=0$, $0.52$, $1.02$ and $2.08$ are shown in Figure
\ref{ps}. At small scales, the pressure power spectra of the non-adiabatic
simulation (dashed lines) fall below the corresponding adiabatic ones (solid
lines). Most of tSZ signals at these scales are contributed by gas in
less  massive  halos. These halos have shallower potential wells and thus SN
feedback is easier to deplete gas. This suppression effect is
even larger toward higher redshifts where the gravitational potential wells
are shallower, leading to as high as $50\%$ reduction at $z=2.08$. The same
feedback redistributes gas and moves the clustering power of the thermal energy
from small scales to large scales. This is likely the reason that we see enhancement
at large scales (Fig. \ref{ps}). However, at these scales, massive halos
contribute most of the signal. Since they have deeper potential wells to fight
against feedback and confine gas, the large scale pressure power spectrum is
less affected.

The gas pressure power spectrum is often expressed by the gas pressure bias
with respect to the matter distribution, defined as
$b_P(k)=\sqrt{\Delta^2_{P}(k)/\Delta^2_{m}(k)}$.  The measured pressure bias
is shown in Figure \ref{bias}. We find in Figure \ref{bias} that the results
are in good agreement with the simulation results at $z=0$ by
\citet{Refregier02}.

\bfi{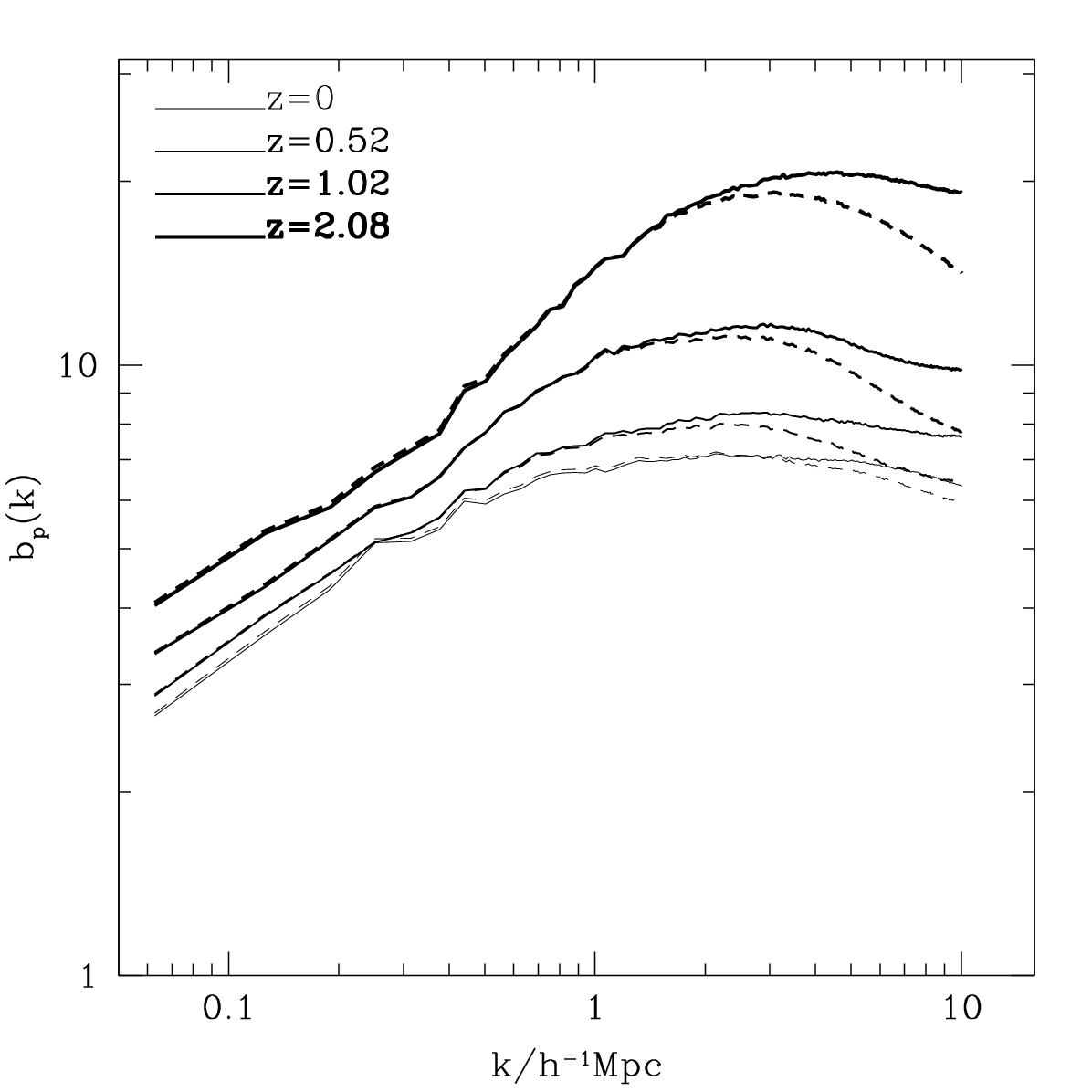}
\caption{The bias $b_P(k)=\sqrt{\Delta^2_P(k)/\Delta^2_m(k)}$ of gas pressure
relative to the dark matter at the four redshifts. The same legends are
adopted as in Figure \ref{tav}.}
\label{bias}
\efi

The SZ power spectrum evaluated from the Limber integral is shown in Figure
\ref{cl}. In the existence of the cooling and self-regulated star formation, the
tSZ angular power spectrum at large scales is suppressed by $\sim 20\%$, which
is mainly due to the reduction of density weighted temperature $T_\rho$. At
the same time, the powers are reduced by as large as around $\sim 40\%$ at
very small scales around $l=10000$, which is jointly caused by the suppression
of both the density weighted temperature and pressure power spectra compared
to the adiabatic simulation.  These suppression effects confirm the previous
results by \citet{daSilva01, White02}. And what's more, the non-adiabatic
power spectrum peaks on a slightly larger scale, about $l\sim 8000$, than the
adiabatic one, which may arise from SN feedback that would drive away the gas
out to diffuse regions.  \citet{SWH01} also found that energy injection would
suppress the power on small scales and push the powers toward larger
scales.  \citet{2007MNRAS.378.1259R} considered the same non-gravitational
processes as our non-adiabatic one in their simulation, and their power
spectrum behaves consistently with ours, despite of their low $y$. 

\bfi{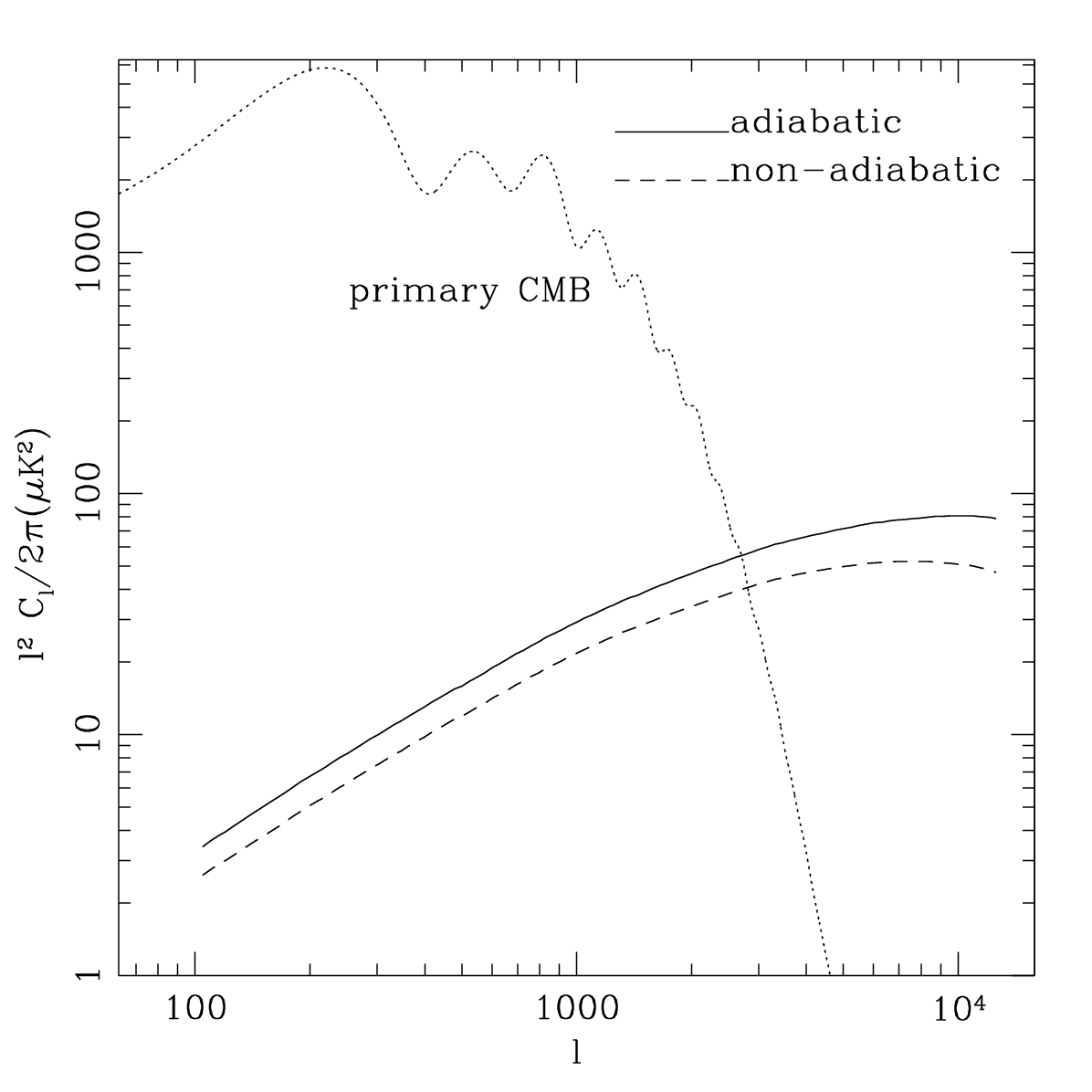}
\caption{The SZ angular power spectra $C_l^{\rm tSZ}$ for the two simulations.
Radiative cooling and SN feedback suppress the SZ power spectrum by $\sim 20\% $
at large scales and around $\sim 40\%$ at small scales. The same legends are
adopted as in Figure \ref{tav}.} 
\label{cl}
\efi

\subsection{The correlation coefficients r(k,z)}
\label{subsec:r}
As addressed in \S \ref{sec:tomography}, a key input of the SZ tomography is
the gas pressure-dark matter density cross correlation coefficient
$r$. State of art numerical simulations with purely gravitational heating
(adiabatic simulations) is able to model $r(k,z)$ robustly. If we can further
quantify the dependence of $r(k,z)$ on additional gastrophysics such as
radiative cooling and feedback, we will be able to model $r(k,z)$ robustly for
general cases and perform the SZ tomography robustly. For this purpose, we
analyze the behavior of $r(k,z)$ in the two simulations.

\bfi{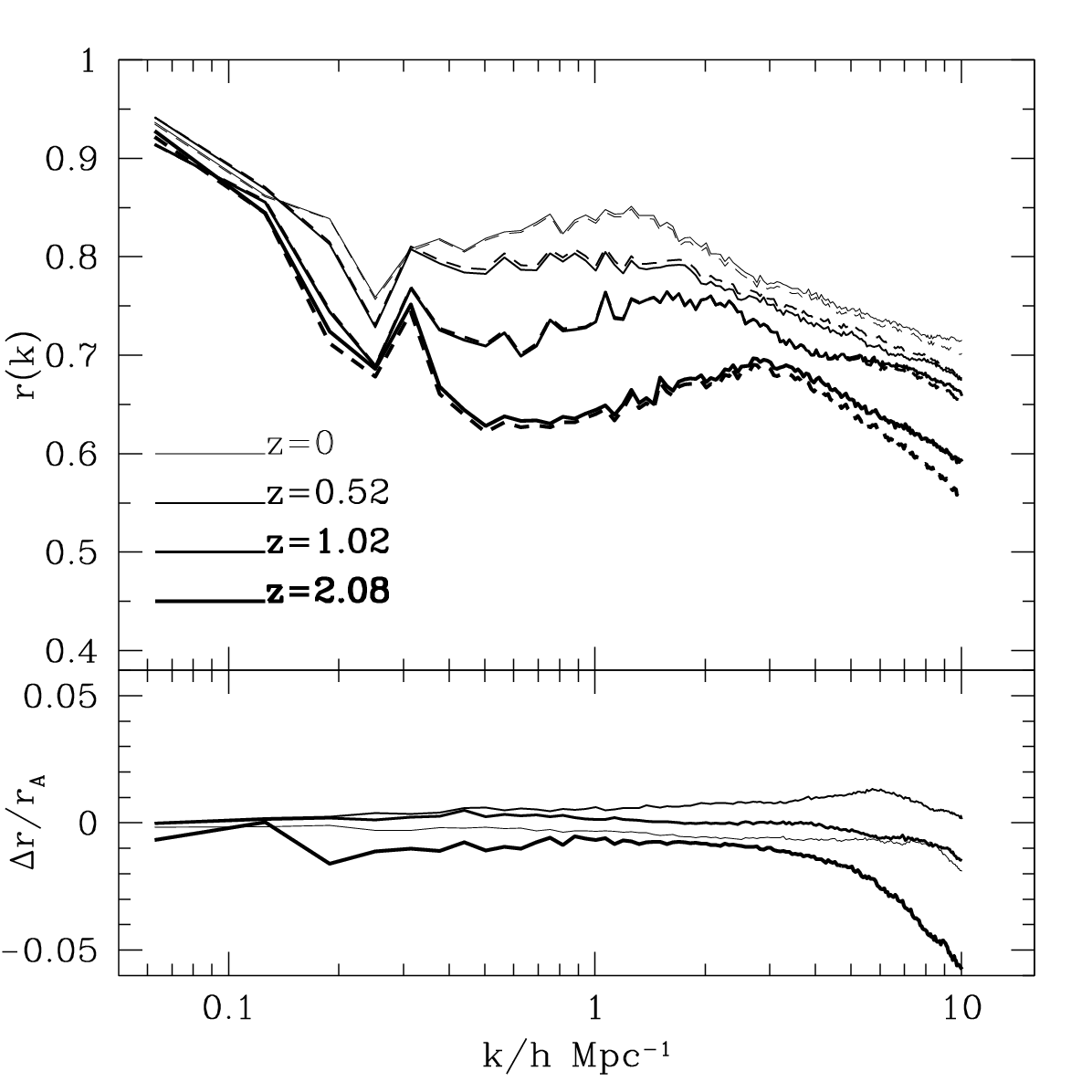}
\caption{\textbf{Top panel}: The cross correlation coefficient between the 3-D gas
pressure and dark matter in adiabatic simulations (solid lines ) and
non-adiabatic simulations (dashed lines). For both the simulations, $r(k)$ is
greater than $0.9$ at $k\simeq 0.062$ \hMpc for all the 4 redshifts, as gas
pressure traces dark matter on the very large scales. At intermediate scales
where $0.4 \leq k \leq 3$ \hMpc, there's still great correlation about $\sim
0.8$ at $z=0$, while $r(k)$ drops by 25\% to $\sim 0.6$ at $z=2.08$.
\textbf{Bottom panel}: The difference of $r$ between the two simulations:
$\Delta r(k)/r_A(k)= (r_{NA}(k)-r_A(k))/r_A(k)$. If cooling and SN feedback
are included, there's only $\sim 1-2\%$ change in $r(k)$ at most scales, and
only at the smallest scales does $r$ change $\sim 5\%$.
}
\label{cc}
\efi
 
The result on $r(k,z)$ is shown in Figure \ref{cc}. We find gas pressure tends
to trace the matter distribution faithfully on the very large scales.  At the
largest scale that our simulation can approach,  $k\simeq 2\pi/L=0.062$ \hMpc,
$r(k)\ge 0.9$ for all of the 4 redshifts.  At $0.4 \leq k \leq 3$ \hMpc,
there's still great correlation between pressure and matter, while $r(k)$
changes by $25\%$ from $\sim 0.8$ at $z=0$ to $ 0.6$ at $z=2.08$. However, at
each redshift, the correlation strength nearly keeps unchanged in this range.
At smaller scales $k \ge 3$ \hMpc, $r(k)$ begins to decrease faster. 

	We further investigate the influences caused by cooling and star
formation on $r(k)$. Surprisingly, we find that these additional gastrophysical
processes have little impact on $r(k)$. In most of the relevant $k$ range, the
change is $\sim 1\%$. Only at very small scales (corresponding to $l\ga 10^4$),
can the effect reach $5\%$ level. This behavior is however explainable, since
the gastrophysics influences the denominator and numerator of $r(k)$ in
basically the same way. A $1\%$ error in $r$ results into a $2\%$ error in the
reconstructed $\Delta^2_PW_{\rm tSZ}^2$, negligible to the $\sim 30\%$ change
in $\Delta^2_PW_{\rm tSZ}^2$ induced by cooling and feedback. Therefore, this
result enables us to extract robust predictions for $r$ from adiabatic
simulations without considering additional complicated gastrophysics, and then
carry out the tSZ tomography method by applying $r$ back into the
observations. Since eventually we need a large sample of simulations with
high resolution to improve the calculation, this feature can significantly save
the computation cost. More importantly, it breaks the otherwise circular
procedure to do the tSZ tomography.\footnote{If $r(k)$ strongly depends on the
detailed gastrophysics, the tSZ tomography procedure will become circular. In
this case, we need to know the detailed gastrophysics to reliably predict
$r(k,z)$ from simulations and then apply back to observations, while we count
on observations to tell us the detailed gastrophysics.} This feature
assures us that we can rely on adiabatic simulations alone to provide $r$.
However, it does not mean that the simulations presented in this paper
have already robustly predicted $r$. We expect the function $r(k)$ to be
sufficiently smooth with respect to $k$. However, since our box size is only
$100h^{-1}$ Mpc, statistical fluctuations induced by cosmic variance are
non-negligible and the simulated $r$-$k$ relation shows clear irregularities
(Fig. \ref{cc}). To beat down the cosmic variance, we need more simulations or
larger box size.  Nonetheless, we only need adiabatic simulations for this
improvement.

\subsection{The feasibility of tSZ tomography in real surveys}
We are then able to evaluate the applicability of the proposed tSZ tomography
to real surveys. As explained in previous sections, it is determined by two
key factors. 
\bi
\item How well are we able to measure the cross correlation (e.g. the
  tSZ-matter correlation or the tSZ-galaxy correlation)?
\item How robustly can we interpret the measured cross correlation and convert
  it into the redshift resolved gas pressure auto-correlation? This last
  quantity is a direct measure of the thermal history.
\ei
Based on the cross correlation coefficient $r$ obtained in \S \ref{subsec:r}
from the simulations, we are able to make predictions for target SZ surveys and
lensing surveys. We show the calculation of the signal to noise ratio of the
tSZ-lensing cross power spectrum in Appendix \S \ref{app:lensing}, and find that
SZ surveys such as SPT and lensing surveys such as LSST allow us to measure the
tSZ-lensing cross correlation to high precision over a dozen source redshift
bins. Refer to the appendix \S \ref{app:lensing} for detailed calculation. Given
such precision measurement and given the robustness of $r$ against cooling and
feedback, we are able to perform the tSZ tomography robustly (\S
\ref{app:lensing}). 

TSZ-galaxy number density  cross correlation can be measured to even higher
accuracy, since galaxy number density measurement is much easier and much more
accurate than weak lensing measurement. So in principle, galaxy surveys, in
combination with SZ surveys,  are also suitable targets to perform the SZ
tomography.  One subtlety is to interpret the measured cross correlation signal.
To be more specific, how well are we able to understand the cross correlation
coefficient between the gas pressure and the galaxy number density, given large
uncertainties in gastrophysical processes such as feedback and cooling.  If the
galaxy bias is deterministic, our study in previous sections has provided robust
answer to this question. However, in reality, the galaxy bias contains some
stochasticity. Further, the amount of stochasticity can be affected by the
presence of feedback and cooling. In  Appendix \S \ref{app:galaxy}, we show that
its impact can be of $\sim 10\%$. This makes  the interpretation of the measured
tSZ-galaxy number density cross correlation challenging.

\section{Conclusion and discussion}
\label{sec:discussion}
In this paper, we investigate the tSZ tomography method,
which aims to extract the redshift information of the tSZ effect. The basic
idea is to cross correlation the tSZ map with tracers of the large scale
structure (galaxy distribution or weak lensing) and recover the redshift
information by the aid of the redshift information of the large scale
structure tracers. The key ingredient which determines the feasibility of
the tSZ tomography is the cross correlation coefficient $r$ between the gas
pressure and the 3D matter (galaxy) distribution. For the tSZ tomography to be
doable, $r$ must be sufficiently large such that the cross correlation can be
measured to reasonable accuracy. To convert the measured cross correlation to
the time resolved tSZ effect, namely the 3D gas pressure auto-correlation power
spectrum, $r$ must be understood robustly, given the presence of gravitational
heating, radiative cooling, star formation and feedback. 

Through our simulations, we find that $r$ is in general sufficiently large
(e.g. larger than $0.5$) over the range of interest. Hence it is in general
feasible to robustly measure the cross correlation between the tSZ effect and
the large scale structure. For example, in the Appendix \S \ref{app:lensing}, we
calculate the signal to noise ratio of the tSZ-shear cross power spectrum
$C_\ell^{\rm tSZ-\gamma}$ for target SZ surveys and lensing surveys, and find
that the cross correlation between the tSZ effect and weak lensing can be
measured with over $100\sigma$ combining SPT and LSST. So the key question in
the tSZ tomography is whether we can robustly predict $r(k,z)$, given
uncertainties in gastrophysical processes. To quantify the effect of two
dominant gastrophysical processes, namely radiative cooling and SN feedback, we
compare the result of our adiabatic simulation against the one with cooling and
feedback. We find that $r(k,z)$ is insensitive to these additional
gastrophysics. The resulting difference in $r(k,z)$ is $\sim 1-2\%$ on most
relevant $k$ and $z$ range, much smaller than the $\sim 30\%$ change in the SZ
power spectrum and the gas density weighted temperature. This allows us to
neglect the dependence of $r(k,z)$ on these gastrophysical processes and adopt
the $r(k,z)$ evaluated from adiabatic simulations as the input of the tSZ
tomography. We thus show that the tSZ tomography with weak lensing surveys is
highly feasible. On the other hand, the galaxy-matter relation and hence $r$ can
be affected by these gastrophysics at $\sim 10\%$ level. These complexities must
be understood first to perform the tSZ tomography with galaxy surveys. 

There are many remaining issues to investigate in future works. (1) The tSZ
tomography relies on the cross correlation measurement between the tSZ signal
and the large scale structure. Hence, any residual noise in the tSZ measurement
associated with the large scale structure, such as residual contamination from
dusty star forming galaxies, could bias the cross correlation measurement.
Multi frequency band information helps to reduce these potential contaminations.
Furthermore, these contaminations could have different redshift dependence from
the tSZ signal and such differences provide another possibility to extract the
tSZ signal. This issue is of crucial importance for further investigation.
(2) Our simulation needs a number of improvements to investigate the
gastrophysics more realistically. We need many more realizations to reduce the
cosmic variance. We may also need to correct for the alias effect to improve the
accuracy of $r$ at scales $k\ga 4 $\hMpc. One gastrophysical mechanism not
included in our simulation is the AGN feedback, whose impact on $r$ and the tSZ
tomography is an important issue for further investigation. (3) A more
comprehensive and quantitative analysis on the performance of the tSZ tomography
with lensing surveys shall be carried out in the future. 

\section{Acknowledgement}
We'd like to thank the anonymous referee for helping improve the paper. We thank
Volker Springel for providing his numerical simulation result for comparison and
Klaus Dolag for useful conversation. The simulations were done at Shanghai
Supercomputer Center by the supports of Chinese National 863 project (grant
No.06AA01A125). This work is supported in part by the National Science
Foundation of China (grant No. 10543004, 10821302, 10873027 \& 10878001), the
Knowledge Innovation Program of CAS (grant No. KJCX2-YW-T05 \& KJCX3-SYW-N2) and
the 973 program grant No. 2007CB815401 \& 2007CB815402.

\appendix
\section{Error forecast on the tSZ-lensing cross correlation measurement}
\label{app:lensing}
In the main context, we have shown that the cross correlation between the gas
pressure and matter distribution is not only strong, but also insensitive to 
extra gastrophysics on scales of interest as well as at redshifts of interest.
It assures in theory the applicability of the proposed tSZ tomography method.
Here, we forecast the S/N when applying this tomography technique to planned SZ
and weak lensing surveys.  

\begin{table}
\begin{center}
\caption{The adopted parameters in the galaxy lensing surveys (DES and LSST)
and the SZ surveys (PLANCK, ACT and SPT)}
\label{tab:surveys}
\begin{tabular}{ccccc}
\hline\rule{0pt}{3ex}
&\multicolumn{2}{c}{$f_{\rm sky}$\footnotemark[1]}&$\sigma_{\rm p,T}$
&$\theta_p$	\\
&DES\footnotemark[2]	&LSST\footnotemark[2]	&$\mu K$	&arcmin	\\
\hline
\rule{0pt}{2ex}
PLANCK 	&0.12 	&0.5 	&4.8 	&5\\
\rule{0pt}{2ex}
ACT	&0.025 	&0.025  &2 	&1.7\\
\rule{0pt}{2ex}
SPT	&0.1 	&0.1 	&2 	&1\\
\hline
\rule{0pt}{2ex}
\end{tabular}
\end{center}
\footnotetext[1]{$f_{\rm sky}$ is the fractional sky coverage of overlapping
  regions between a given SZ survey and a given lensing survey.}
\footnotetext[2]{The galaxy number density per steradian $\bar{n}_g=1.2\times
10^7 \bar{n}^{\prime}_g$. $\bar{n}^{\prime}_g$ is the galaxy number density per
square arcmin,  $15$ for DES and $40$ for LSST.} 
\end{table}

As stated by \citet{Massey07}, the matter distribution can be extracted
from the lensing tomography technique. So, in an equivalent way, the S/N of the
tSZ tomography with lensing surveys can be demonstrated by the S/N of the cross
power spectrum $C_l^{\rm tSZ-\gamma}$, where the source galaxies of lensing
measurement locate between a thin redshift bin (e.g. $\Delta z=0.2$). The
measurement error is
\begin{equation}
\label{eqn:signaltonoise}
 \left(\frac{S}{N}\right)^2=\sum_l^{l_{\rm max}} \frac{2l\Delta l f_{\rm
sky}\left(C_l^{\rm tSZ-\gamma}\right)^2}
{\left(C_l^{\rm
tSZ-\gamma}\right)^2+\left(C_l^{\rm tSZ}+C_l^{\rm CMB}+C_l^{\rm det}+C_l^{\rm
DSFG}\right) \left(C_l^{\gamma}+C_l^{\gamma,N}\right)}\ .
\end{equation}
Here, $f_{\rm sky}$ is the combined sky fraction between lensing survey and
SZ survey. The tSZ power spectrum $C_l^{\rm tSZ}$, the lensing power spectrum
$C_l^{\gamma}$ and the cross power spectrum $C_l^{\rm tSZ-\gamma}$ are all
calculated from simulations. To better account for the real surveys, we consider
the non-adiabatic simulation and at the frequency 150 GHz which ACT and SPT both
cover. Considering the frequency dependence $g(x)\sim-0.95$ at 150 GHz instead
of $g(x)\sim -2$ in RJ band used in the main context, the predicted tSZ angular
power spectrum $C_l^{\rm tSZ}$ should be brought down by 4.43 times and
$C_l^{\rm tSZ-\gamma}$ down by 2.1 times.  We adopt $l_{\rm max}=6600$ in
  the numerical estimation.  Around this scale, both the tSZ and the lensing
  measurements are already noise dominated. So a cut at $l=6600$ does not
  cause significant underestimation of S/N. 

 The lensed primary CMB power spectrum $C_l^{\rm CMB}$ is calculated from
CAMB\footnote{http://camb.info}\citep{Lewis2000}. The beam of CMB
  experiment affects both the cross correlation signal and the statistical
  error. However, in the S/N estimation, its effect can be compressed into a
  single term $C_l^{\rm det}$, 
\begin{equation}
\label{eqn:detnoise}
 C_l^{\rm det}=(\sigma_{\rm p,T}\theta_p)^2
B^{-2}(l)=(\sigma_{\rm p,T}\theta_P)^2\exp(l^2/l^2_{\rm beam}) \ .
\end{equation} 
Here we have approximated the beam $B(l)$ as a Gaussian form with
$B(l)=\exp({-l^2/2l_{\rm beam}^2})$  and $l_{\rm beam}=\sqrt{8
\ln{2}}/\theta_p$.  $\theta_p$ is the pixel size (angular resolution) and
$\sigma_{\rm p,T}$ is the r.m.s instrument noise per pixel. Dusty star forming
galaxies (DSFGs) are found to contaminate the SZ observation significantly
\citep{Fowler2010, Hall2010}, so we include this term in the noise estimation.
For the DSFG contamination $C_l^{\rm DSFG}$, we adopt  Poisson distribution
$l^2 C_l^{\rm DSFG}/2\pi=10 (l/3000)^2 (\mu K)^2$ suggested by recent ACT
and SPT results. $C_l^{\gamma,N}$ is the power spectrum
of the shape error due to the intrinsic ellipticities of source galaxies. The
redshift distribution of source galaxies we adopt in lensing surveys follows
\citet{Huterer2002}
\begin{equation}
 p_s(z)=\bar{n}_g \frac{z^2}{2 z_0^3}e^{-z/z0}\ ,
\end{equation}
which satisfies $\int_0^\infty dz p_s(z)=\bar{n}_g$. Here, $\bar{n}_g$ is the
average number density of galaxies per unit steradian. The peak of the
distribution function is at $2z_0$. We adopt $z_0=0.3$ for DES survey, while
$z_0=0.4$ for LSST survey. For the sake of tomography, source galaxies in the
redshift range $z_s=0.2-1.6$ are divided into several redshift bins with width
$\Delta z=0.2$ (see Table \ref{tab:sncross} for the redshift bins). The galaxy
number density in the redshift bin $z_j-\Delta z/2<z_s<z_j+\Delta
z/2$\footnote{Note here $z_s$ should be higher than the redshift bin of the
contributing dark matter $z_i$, and thus we assume $z_j>z_i$ to avoid
confusion.} is
\begin{equation}
 n_s=\int_{z_j-\Delta z/2}^{z_j+\Delta z/2} dz p_s(z)\ .
\end{equation}
Thus, the galaxy shape noise \citep{Kaiser1998} in this redshift bin reads
$C_l^{\gamma, N}=\gamma^2_{\rm rms}/n_s$, where $\gamma^2_{\rm rms}=0.1$ is
adopted for both lensing surveys. All the adopted parameters are listed in
Table \ref{tab:surveys}.

Note that in Eq. \ref{eqn:signaltonoise}, the SZ power spectrum is integrated
over the entire redshift range, while the lensing power spectrum is integrated up
to the redshift of the source galaxies $z_s$. We show in Table \ref{tab:sncross}
the S/Ns of the cross power spectra between lensing surveys and SZ surveys in
several redshift bins. (1) It turns out that, any pair between one of the three
CMB surveys (ACT, Planck and SPT) and one of the two lensing surveys (DES and
LSST) can measure the tSZ-lensing cross correlation robustly, with an overall
S/N $\ga 20$. By no surprise, SPT+LSST  will enable the highest precision
measurement, with an overall S/N $\ga 200$. (2) However to perform the lensing
and tSZ tomography, Planck+DES may not be suitable, due to relatively low S/N
in each redshift bin (table \ref{tab:sncross}). But at least for SPT+LSST,
the lensing and tSZ tomography is promising. 

We caution the readers that the S/N presented here is overestimated to
  some extent, for two major reasons. First, the estimation is done assuming
  Gaussianity. It is well known that both the lensing and the SZ effect are
  non-Gaussian at scales of interest. Non-Gaussianity boosts the cosmic
  variance and induces correlations between different multipole modes. Second,
  we have neglected potential systematical errors in the cross correlation
  measurement. The distribution of DSFGs is strongly coupled to the large
  scale structure.  So DSFG contamination in the SZ maps is correlated with
  galaxies used to perform the thermal SZ tomography. Hence DSFGs 
  contribute not only statistical errors shown in Eq. \ref{eqn:signaltonoise},
  but also systematical error in the cross correlation measurement. This
  systematical error can be removed by the  different spectral dependences of DSFGs
  and the thermal SZ effect, at the expense of loss in S/N.  More realistic
  evaluation of the thermal SZ tomography S/N shall take these complexities
  into account. 

\begin{table*}\begin{center}
\caption{The S/N of the cross power spectra predicted from DES, LSST combined
with PLANCK, ACT and SPT.}
\label{tab:sncross}
\begin{tabular}{c|*{3}{c}|*{3}{c}|}
lensing survey	& \multicolumn{3}{c}{DES} \vline& \multicolumn{3}{c}{LSST}\vline
\\
\hline
\rule{0pt}{3ex}
SZ surveys& PLANCK	&ACT 	&SPT &PLANCK	& ACT 	& SPT	\\
\hline
\rule{0pt}{3ex}
$z_s=1.4-1.6$   &9.90  &25.34  &65.37  &33.97  &53.13  &137.58	\\
\rule{0pt}{2ex}
$z_s=1.2-1.4$	&10.81 &26.69  &68.38  &34.25  &51.55  &132.58	\\
\rule{0pt}{2ex}
$z_s=1.0-1.2$  	&11.25 &26.34  &66.85  &33.09  &47.01  &119.75	\\
\rule{0pt}{2ex}
$z_s=0.8-1.0$	&10.92 &23.70  &59.40  &30.04  &39.20  &98.54 	\\
\rule{0pt}{2ex}
$z_s=0.6-0.8$	&9.54  &18.59  &45.80  &24.67  &28.55  &70.49 	\\
\rule{0pt}{2ex}
$z_s=0.4-0.6$	&6.93  &11.62  &27.97  &16.88  &16.57  &39.92 	\\
\rule{0pt}{2ex}
$z_s=0.2-0.4$	&3.37  & 4.52  &10.52  & 7.69  & 5.98  &13.92  	\\
\end{tabular}\end{center}\end{table*}

\section{Complexities to perform the tSZ tomography with galaxies}
\label{app:galaxy}
Galaxy redshift survey is an obvious alternative to perform the tSZ tomography.
Indeed, the tSZ-galaxy (number density) cross correlation can be measured with
higher S/N than the tSZ-lensing cross correlation measurement.  However,
uncertainties exist in interpreting the measured cross-correlation. The measured
cross-correlation signal is proportional to the cross correlation coefficient
$r_{Pg}$ between the gas pressure and the galaxy number density, while this
quantity suffers from uncertainties in gastrophysics. We have shown that
$r_{Pm}$ is insensitive to gastrophysics. Then the key is the dependence of
$r_{mg}$, the cross correlation coefficient between matter and galaxy
distribution, on gastrophysics. At sufficiently large scales, $r_{mg}=1$.
However, at scales of interest, stochasticities in galaxy distribution cause
$r_{mg}\neq 1$ (e.g. \citealt{Bonoli09,Baldauf2010}). Gastrophysics induces
extra stochasticities in galaxy distribution and hence affects $r_{mg}$.

We are at no position to develop a realistic galaxy formation model and robustly
quantify $r_{mg}$ and its gastrophysical dependence. However, the halo occupation
distribution (HOD) model \citep{CS2002} provides us a convenient tool to check
for possible dependences. Due to uncertainties in the adopted HOD parameters,
results presented here are inconclusive, and only serve to demonstrate
uncertainties in $r_{mg}$ and hence in interpreting the measured tSZ-galaxy
cross correlation. 
\begin{figure}[htb!]
\centering%
\includegraphics[width=4in]{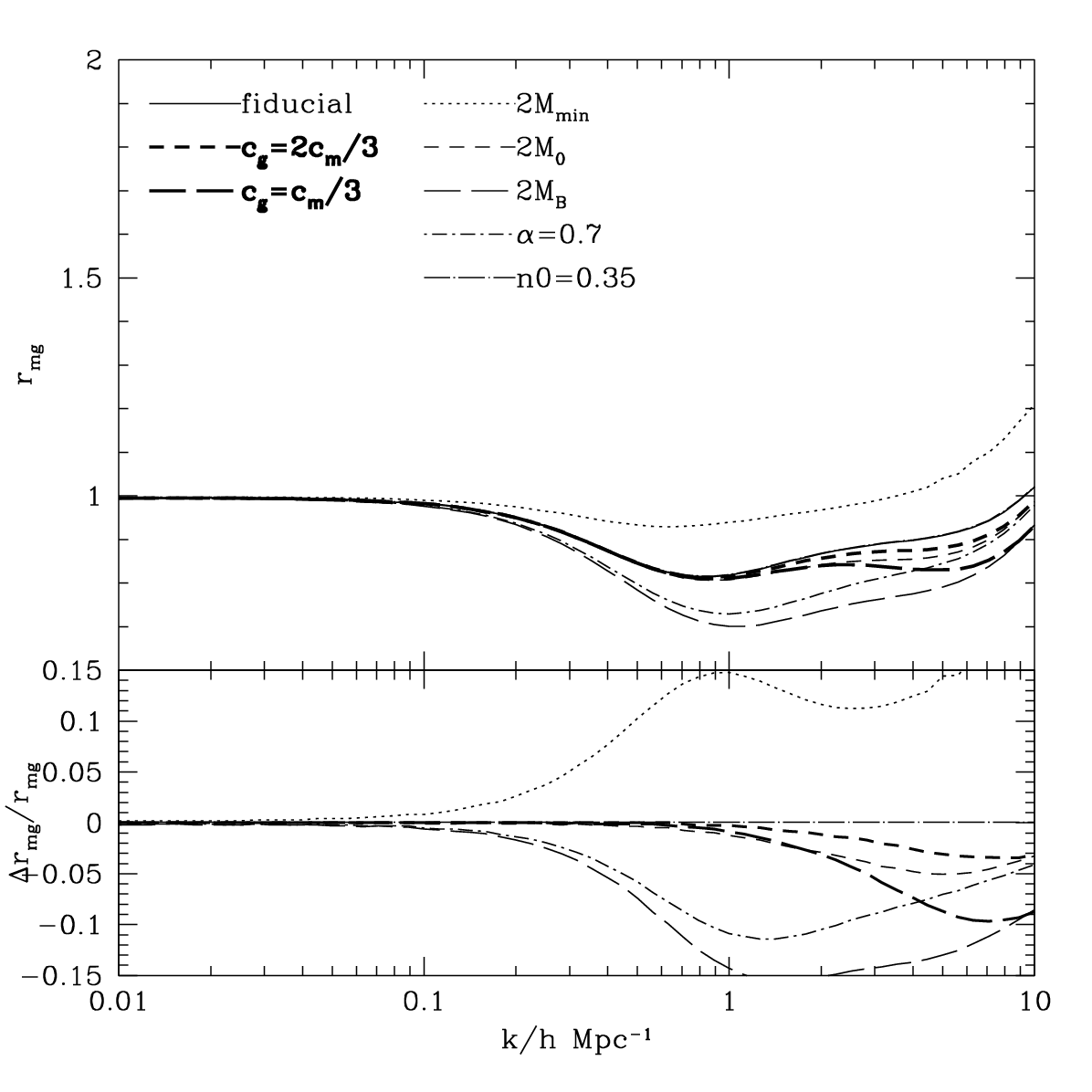}
\caption{Dependences of the dark matter-galaxy cross correlation
coefficient $r_{\rm mg}$ on  HOD parameters.  These dependences imply a likely
non-negligible dependence of $r_{mg}$ on gastrophysics, which leads to changes
in HOD parameters. }
\label{fig:rhod}
\end{figure}

The adopted HOD model has six parameters to describe the galaxy distribution
within a halo. (1) The mean galaxy number $\left<N_{\rm
gal}|M\right>$ in a halo of mass $M$ is \citep{Sheth2001,Cooray2004}, 
\begin{eqnarray}
	\left< N_{\rm gal}|M \right> = \left\{ \begin{array}{rcl}
	N_0, & \mbox{ for } &  M_{\rm min} \le M \le M_B \\
	N_0 (M/M_B)^{\alpha}, & \mbox{ for } & M > M_B
	\end{array}\right.  \ .
\end{eqnarray}
(2) The galaxy number distribution is described by the binomial distribution
\citep{Scoccimarro2001} with
\begin{equation}
	\left< N_{\rm gal}(N_{\rm gal}-1)|M \right>^{1/2}=\beta(M) \left< N_{\rm
gal}|M \right>,
\end{equation}
where $\beta(M)=\log{\sqrt{M/M_{\rm min}}}$ for $M<M_0$, and $\beta(M)=1$
otherwise. (3) The galaxy space distribution within the halo is described by
the NFW \citep{NFW1997} profile, quantified with a galaxy concentration
parameter $c_g(M)$. Notice that it is not necessarily equal to the concentration
parameter of the matter density, $c_m(M)$, for which we adopt
$c_m(M)=9(M/M_{\star})^{-0.13}$ \citep{Bullock2001}. For example,
\citet{Yang2005} and \citet{Nagai2005} suggest $c_g\simeq c_m/3$.

The fiducial model has 
\be
\left(M_{\rm min},M_B, N_0,\alpha, M_0, c_g(M)\right)=\left(10^{11}
h^{-1} M_\odot, 4\times 10^{12}h^{-1} M_\odot, 0.7, 0.8, 10^{13}h^{-1}
M_\odot,c_m(M) \right)\ .
\ee 
Gastrophysics may be mimicked by varying these parameters. For example, feedback
could suppress galaxy formation, resulting in smaller $N_0$ or larger $M_{\rm
min}$. It can also make the galaxy spatial distribution less clumpy than the
dark matter distribution, resulting in $c_g<c_m$. We show a number of examples
in Fig. \ref{fig:rhod}. In general, varying these parameters can affect $r_{mg}$
at $10\%$ level and implies that gastrophysics can introduce similar
uncertainty. Hence we have to keep at caution to interpret the measured
tSZ-galaxy cross correlation. 

\end{document}